# Conducting more inclusive solar geoengineering research: A feminist science framework

**Short title: Inclusive geoengineering research framework**

**Ben Kravitz[1,2,*] and Tina Sikka[3]**

[1]Department of Earth and Atmospheric Sciences, Indiana University, Bloomington, IN, USA.

[2]Atmospheric Sciences and Global Change Division, Pacific Northwest National Laboratory, Richland, WA, USA.

[3]Department of Media, Culture, and Heritage, School of Arts and Cultures, Newcastle University, Newcastle, UK.

[*]Corresponding author: Ben Kravitz (bkravitz@iu.edu)

**Abstract**

Solar geoengineering, or deliberate climate modification, has been receiving increased attention in recent years. Given the far-reaching consequences of any potential solar geoengineering deployments, it is prudent to identify inherent biases, blind spots, and other potential issues at all stages of the research process. Here we articulate a feminist science-based framework to concretely describe how solar geoengineering researchers can be more inclusive of different perspectives, in the process illuminating potential implicit bias and enhancing the conclusions that can be gained from their studies. Importantly, this framework is an adoptable method of practice that can be refined, with the aim of conducting better research in solar geoengineering. As an illustration, we retrospectively apply this framework to a well-read solar geoengineering study, improving transparency by revealing its implicit values, conclusions made from its evidence base, and the methodologies that study pursues. We conclude with a set of recommendations for the geoengineering research community whereby more inclusive research can become a regular part of practice. Throughout this process, we illustrate how feminist science scholars can use this approach to study climate modeling.

**Keywords**

This manuscript was borne out of a period of reflection for both of us. Around the same time, we separately published blog posts taking a retrospective look at the field of geoengineering research (see below). We noticed several critical gaps, as well as potential ways of bridging these gaps and improving research practices. A team-up seemed obviously fruitful, the result of which is this manuscript.

Here we introduce a framework, based on Feminist Contextual Empiricism, that is aimed at introducing pluralism to geoengineering research and providing a systematic way of laying bare many assumptions in the research process. We retrospectively apply this framework to a well-read solar geoengineering climate modeling study. In doing so, we illustrate how feminist inquiry can be an adoptable method of practice for solar geoengineering researchers to avoid blind spots, include a more holistic understanding of results and conclusions, and ultimately result in better science. In addition, through this framework, we illustrate a potential path for feminist science scholars to apply their methodologies to climate modeling. Through synergies between these two different disciplines, we aim to identify the added value these two disciplines can provide to each other.

Interdisciplinary long-form discussion manuscripts do not have an obvious home. We considered more than 30 journals to which we did not submit because they were poor fits for the piece we had written. We did submit to seven different journals and were desk rejected from all of them: not what the journal was looking for at this time, not focused enough on research in one specific discipline, or an inability to find qualified reviewers. After eight months since our first submission, not a single peer reviewer has evaluated our manuscript.

Regardless of our frustrations with the review process, we decided that further delaying this manuscript (or risking it never seeing the light of day) was not something we wanted to do. Hence, we posted it on the arXiv. We're not claiming that it's perfect, but we hope that you find it useful and that it stimulates conversation.

Sincerely,

Ben Kravitz and Tina Sikka

Blog posts
https://geoengineering.environment.harvard.edu/blog/ten-years-geomip
https://www.c2g2.net/gender-and-climate-engineering-a-view-from-feminist-science/

# 1 Introduction

Solar geoengineering, or the deliberate modification of the climate by reducing solar energy at Earth's surface, has received increased attention in recent decades as a means of counteracting climate change. These ideas, such as creating a reflective layer of stratospheric aerosols (often sulfate) or brightening low clouds over the ocean, can temporarily offset anthropogenic climate change, allowing for the ramp-up of greenhouse gas emissions mitigation or negative emissions. Scientific research on geoengineering has grown substantially over the past decades (Crutzen, 2006; Oldham et al., 2014), including several national and international assessments of geoengineering (Committee on Developing a Research Agenda and Research Governance Approaches for Climate Intervention Strategies that Reflect Sunlight to Cool Earth et al., 2021; IPCC, 2013; National Research Council, 2015a, 2015b; Shepherd et al., 2009).

Most of the research thus far in solar geoengineering has been directed by researchers in rich, industrialized countries in the Global North (Biermann & Möller, 2019). While these studies have been instrumental for advancing scientific understandings of solar geoengineering and its risks, the predominance of certain perspectives in producing knowledge and discourse can lead to blind spots and other serious policy and communication problems (DeLoughrey et al., 2015; J. A. Nelson, 2008; Sikka, 2018). In climate engineering and negative emissions technologies, narrow perspectives have led to mistaking modeling feasibility for real-world feasibility (Low & Schäfer, 2020); normalizing particular topics of discourse, which shapes policy (Beck & Mahony, 2018) or de facto governance (Gupta & Möller, 2019); and the enshrinement of particular values that are then imposed more broadly (Oomen, 2019). Because there have historically been few practical avenues for researchers from the Global South to join the geoengineering research community (Winickoff et al., 2015), leading to numerous debates in solar geoengineering meetings about what the developing world thinks (sometimes even phrased so reductively as to lump the entire Global South into a single entity). Recognition of these issues (Buck et al., 2014; McLaren, 2018) has led to more proactive efforts to include developing country perspectives and build indigenous research capacity in developing countries (Rahman et al., 2018). Nevertheless, it is likely that some perspectives will play a prominent role in solar geoengineering research for the foreseeable future. In a topic like solar geoengineering, which literally involves modifying Earth's climate, the humility to recognize and address blind spots is of paramount importance.

There are concrete steps that solar geoengineering researchers can take to include diverse perspectives, with tangible positive outcomes for both the science and the researchers. A feminist science approach embodies these values of inclusivity and multiple perspectives, providing the foundation for a framework whereby scientists can better interpret their results for more policy-relevant conclusions and include broader perspectives that challenge implicit biases and black-boxed assumptions (Sikka, 2018). By adopting this framework, solar geoengineering research can engage in better science, wherein "better" means facilitating more inclusive, representative, reflexive, and pluralistic scientific practice and outcomes. In addition, through an examination of solar geoengineering, we illustrate an approach whereby feminist science scholars can apply their methods to climate modeling. Through these synergistic approaches, we aim to demonstrate the added value these two different disciplines can provide to each other.

## 2 Feminist Science as a Practice

## 2.1  A brief introduction to Feminist Contextual Empiricism

A goal of feminist science is to open up new avenues of thinking and analysis while attending to gender and other forms of bias by building on the social study of science (Pickering, 1992). Feminist science focuses specifically on issues of representation and diversity in the sciences, serving as an adoptable method of practice by providing tools to increase inclusivity.

We apply Feminist Contextual Empiricism (FCE) (Longino, 1990, 2019; Longino & Lennon, 1997), which is rooted in the premise that science is value-laden (wherein objectivity and impartiality are established norms) and argues that an important goal of science is to pursue better values.  FCE expands upon the values of accuracy, consistency, broad scope, simplicity, and fruitfulness (Kuhn, 1977) to embrace equity and justice wherein truth is achieved through discursive consensus formation.  By explicitly including multiple interpretations of the same phenomena, values and underlying assumptions are laid bare, revealing the evidence bases for arguments and theories (Longino, 1990).  Although the original iteration of FCE focused on gender differences in experiential knowledge and the unequal power relations in which women are enmeshed, it has since grown (in some cases converging with other avenues of thought) to encompass additional feminist values including heterogeneity and mutuality of interaction in which plurality and complexity of explanation are prioritized. FCE also contends that research must be applicable to human needs, for example, aiming to alleviate misery (Longino, 1996). These norms do not constitute a fundamentally new approach to science, but rather an evolution that is consistent with the idea of scientific revolution (Longino, 1987).

At the root of FCE is the accumulated experiential knowledge of women stemming from their engagements with unequal relations of power (both interpersonal and structural). The biases women face are persistent, pervasive, and in many cases can be overtly hostile (Committee on Increasing the Number of Women in Science, Technology, Engineering, Mathematics, and Medicine (STEMM) et al., 2020).  The understanding that emerges out of these experiences can be used as a resource.  Similar examples of power disparities that commonly occur in research environments include those of race, nationality, wealth, sexuality, career stage, and geography.  In science, it is often the case that the perspectives of the marginalized are repressed in favor of the dominant group, which can stifle creativity, diminish the agency of the workforce and, in some cases, result in the persistence of harmful lines of thought (Lloyd, 2009; Metoyer & Rust, 2011; Young et al., 2019).  Rather than enduring the dominance of one kind experience or background or, in extreme cases, the assertion that one set of views or conclusions is "right", FCE provides a pathway for incorporating pluralism, including multiple sources of hypotheses and analyses into research resulting in a more holistic understanding of the problem at hand (Wylie, 2007).  This is what makes feminist empiricism not *feminine* (thereby eschewing gender essentialism), but *feminist.* FCE is fundamentally about examining how scientists practice science, specifically how research is conducted in a way that is consistent with the scientists' values (Longino, 1987, 1990).  In addition to a practice by which science is performed, FCE (and frameworks like it that focus on inclusion of a variety of perspectives) can be used as a tool to reevaluate previous studies or conclusions to obtain stronger or more actionable results.

In addition, feminist science contends that a critical aspect of feminist critique involves the re-examination of biased language (e.g., metaphors and descriptions) that shape scientific practice. Scientific practice is often a reflection of society, including societal gender bias, and in turn science can serve as justification for gender discrimination (Sheets, 2003), including in solar geoengineering (Buck et al., 2014; Sax, 2019).

## 2.2 Feminist Retrospective Analysis

In addition to a practice by which science is performed, FCE (and frameworks like it) can be used as a tool to reevaluate previous studies or conclusions. The following are three examples of feminist retrospective analysis: human fertilization, machine learning, and laboratory science. These examples help illustrate how biased viewpoints have resulted consequential erroneous conclusions, as well as how more inclusive practices can help avoid these pitfalls.

1. Although fertilization is a cooperative process (Nettleton, 2015; Trogen, 2016), many scientific and medical textbooks continue to masculinize the sperm as adventurous, active, and agential and femininize the egg as passive, delicate, and static because these descriptions fit with settled social assumptions about gender roles (Keller, 2002; Martin, 1991). Additionally, a persistent fallacy based on faulty assumptions about scientific complexity, genes, gender, and sex has led to the widespread belief that there is no genetic component to ovary formation (Gilbert & Rader, 1998). These mischaracterizations of biology have led to the neglect of women's reproductive health as evidenced by comparatively fewer studies of female sexuality and pleasure (Lloyd, 2009) as well as inadequate research on medical conditions like endometriosis, fibroids, and polycystic ovarian syndrome; these conditions were believed to be a function of "hysteria" or feminine attention seeking (Metoyer & Rust, 2011; Young et al., 2019).

2. Wu & Zhang (2016) used a seemingly objective machine learning-based face recognition algorithm to study criminality. Their algorithm, which solely evaluates facial features, was up to 89.9% accurate in identifying people with criminal records. However, their algorithm did not incorporate structural biases, such as how economically disadvantaged minorities are underrepresented in facial recognition software training data (Simonite, 2019) or that economically disadvantaged minorities are disproportionately incarcerated (Campbell et al., 2015).

3. The Civic Laboratory for Environmental Action Research (CLEAR) is a self-identified anti-colonial and feminist lab monitoring plastic pollution in Newfoundland, Canada that engages in research grounded in feminist theory to create more opportunities for better science (Rivers, 2019). Their research on plastic contaminants in fish utilizes the help of women and fishermen working on the wharves, as well as members of the general public interested in their "collecting guts for science" initiative (Liboiron et al., 2020). By incorporating perspectives and participants that are not normally present in science, they have been able to expand upon natural science research (e.g., they found that 85% of shoreline waste was plastic, that much of the contaminants came from regional sources, and that smaller beaches had more plastic accumulation) and make science-based policy recommendations (e.g., that moratoria were effective in decreasing macroplastics and could thus could be relied on as a model for microplastics). These
proposals were supported by local communities who were directly involved in the
research process.

Each of these cases indicates the importance of creating space for a transformation of the values
that underpin science, particularly those that perpetuate gender stereotypes through dominating
metaphors, and abstract over place-based knowledge. Hard science also reflects the social
norms and values that permeate society.

## 2.3  A Feminist Science-Based Framework for Conducting and Evaluating Science Research

Building on the principles of FCE, we aim to establish a framework whereby solar
geoengineering research can adopt feminist science values, asking questions like:

1.  Are our chosen models and laboratory practices hierarchical or interactional (and why
    have we decided to consistently favor hierarchy)?
2.  What criteria are we using to justify acceptance of evidence?
3.  Are there alternate explanations or hypotheses that can explain the phenomena under
    investigation?
4.  What are the values that underpin scientific knowledge?
5.  Whose interests do the phenomena we are studying serve?
6.  How do we describe science (e.g., by using gendered language)?
7.  And for what ends are we practicing science?

Scientific knowledge is produced by peoples and cultures but retains a claim to objectivity,
which has led to politicization and blind spots (Bijker, 2017; Hoppe, 2005). Paralleling
movements in other parts of social science, these inquiries aim to interrogate underlying
assumptions.  In climate science, the dominance of specific tools like Earth System Models,
which have their own uncertainties, biases, and assumptions, has led to narrow representations of
complexity and partial conclusions (Schneider, 1997; Shackley et al., 1998).  In addition, the
IPCC process, which aims for consensus, has been criticized for downplaying uncertainty and
dissent and producing overconfidence in conclusions (van der Sluijs, 2012).  There are parallel
arguments for pluralism in other sectors, such as energy, energy substitution, and climate
research, for similar reasons (Sovacool et al., 2020).

## 3  Inclusive Solar Geoengineering Research

Solar geoengineering would not be able to perfectly offset climate change from increased
atmospheric carbon dioxide (Moreno-Cruz et al., 2012).  It would be effective at offsetting a
large portion of the changes in temperature and precipitation (Irvine et al., 2019).  A moderate
amount of geoengineering (Keith, 2013) in a high $CO_2$ future would likely result in a climate
closer to the present day than a future with high $CO_2$ alone for a variety of climate variables
(Tilmes et al., 2013).  We hereafter restrict discussions to solar dimming and stratospheric
aerosol geoengineering.  While other methods of solar geoengineering, like marine cloud
brightening or cirrus thinning, may have applicability to the arguments presented here, we did
not give sufficient attention to these proposed technologies to justify making broad conclusions.

## 3.1 Retrospective analysis of a solar geoengineering study

As an illustration of how this feminist science-based framework might improve or reevaluate previous studies, we revisit "A multi-model assessment of regional climate disparities caused by solar geoengineering" by Kravitz et al. (Kravitz et al., 2014). This study describes analyses of 12 models participating in Geoengineering Model Intercomparison Project (GeoMIP) experiment G1, an idealized solar geoengineering simulation involving an abrupt increase in the $CO_2$ concentration and an abrupt decrease in solar input (Kravitz et al., 2011). The authors then evaluated changes in temperature and precipitation in 22 "Giorgi regions" (Figure 1) covering the populated continents (Giorgi & Francisco, 2000). They did so by linearly scaling the climate model output from GeoMIP to determine what each model says is the right "amount" of geoengineering to offset temperature change, precipitation change, or any weighted combination of the two for any of the regions, as well as the global average (Figure 1). The optimal amount of solar geoengineering minimizes a "dis-utility" (damage) function $D_i$ for each Giorgi region $i$, defined as

$$D_i(w;g) = \sqrt{(1-w)[\Delta T(g)]^2 + w[P(g)]^2} \qquad (1)$$

where $w$ is a weighting function between temperature and precipitation (varies between 0 and 1), and $g$ is the amount of solar geoengineering, where $g = 1$ is defined as the amount that exactly offsets global mean temperature change due to high $CO_2$. $\Delta T$ and $\Delta P$ are the (normalized) amount of temperature and precipitation change, respectively, from the baseline. Positive and negative changes are treated symmetrically.

Their conclusions can be roughly summarized:

- All models show that a sizable amount of solar reduction would bring $CO_2$-caused temperature change closer to the baseline in all 22 regions. Beyond that amount, there is at least one region that is "overcooled" – the amount of temperature change is greater under geoengineering than under climate change.
- In every model, any amount of solar reduction exacerbates the precipitation changes due to climate change in at least one region. That region differs between different models.
- Most combinations of temperature and precipitation (most values of $w$) have the same conclusions as for temperature alone ($w = 0$). This is because all changes were normalized by their respective standard deviations so that temperature and precipitation could be compared directly. Precipitation is highly variable, whereas temperature is comparatively less variable, so even small temperature changes disproportionately affect the metric $D_i$.

Citations to Kravitz et al. (2014) have been used for statements with a variety of positions about differential regional impacts from solar geoengineering: a moderate amount of solar geoengineering will benefit everyone (Winickoff et al., 2015); solar geoengineering would have a mixed effect on precipitation changes, leading to winners and losers (Keith & Irvine, 2016); and a combination of the two positions (Heutel et al., 2016). Each of these statements is consistent with the current state of knowledge, but with different sociopolitical implications and different influences on discourse.

## 3.2 Revisiting Kravitz et al. through an inclusive framework based on feminist science

Here we address (a) how Kravitz et al. (2014) explicitly or implicitly provided answers to the questions in Section 2.3; (b) how their study exemplified inclusive principles or missed opportunities to incorporate inclusivity (and what that inclusivity could have added to the study); and (c) how feminist practices can be used to improve the way their study might have been conducted.

### 3.2.1 Are the chosen models hierarchical or interactional?

By hierarchical modeling, we mean approaches in which a single model (or class of models) is run with a set of inputs provided by another model or data source, and the set of outputs is then handed to another set of models or stakeholders (e.g., impacts models). In contrast, interactional modeling involves coupling multiple classes of models to incorporate feedbacks between them, as is common practice in Detailed Process Integrated Assessment Models (Weyant, 2017).

Kravitz et al. exclusively conducted hierarchical modeling without incorporation of climate-society feedbacks (Calvin & Bond-Lamberty, 2018). This is not a criticism, but rather a reflection of their intent: experiment G1 is highly idealized and, while such idealized simulations are essential for gaining critical knowledge about the functioning of the climate system (Kravitz et al., 2013), they are poorly suited to evaluate impacts, policy, or societal feedbacks (Kravitz et al., 2020).

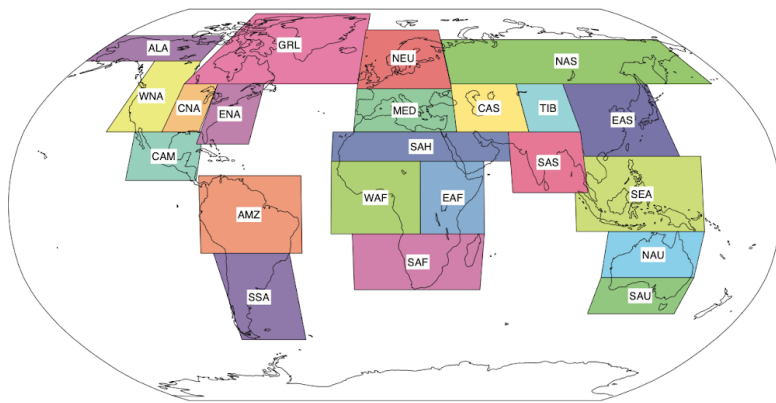

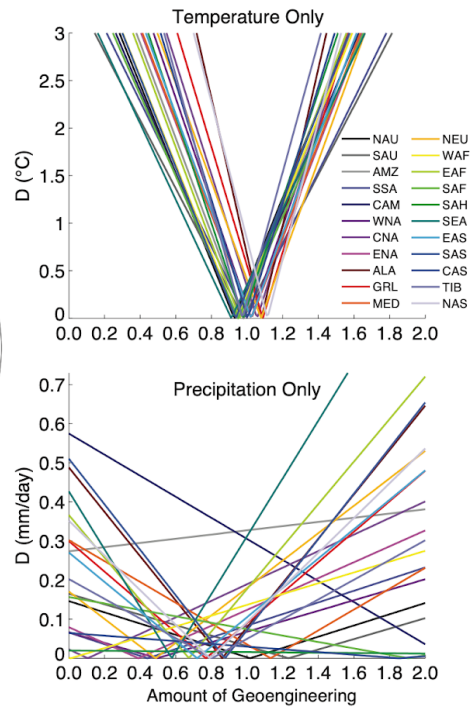

**Figure 1.** The 22 Giorgi regions (left) explored in this study and the results of the 12-model mean for temperature (top right; °C) and precipitation (bottom right; mm/day) for each of those regions. x-axis values of 0 indicate temperature or precipitation changes under a high $CO_2$ world with no geoengineering, and values of 1.0 indicate offsetting global mean temperature change from $CO_2$ with solar geoengineering (solar reduction). y-axes indicate a "dis-utility" function, measuring departures from a preindustrial baseline (values closer to 0 are "better"). All panels are reprinted from (Kravitz et al., 2014) under a CC BY 3.0 license.

Figures 2 and 3 show annual and seasonal mean temperature and precipitation results for a high $CO_2$ world (abrupt4xCO2) and the high $CO_2$ world with solar reduction (G1). As Kravitz et al. state, the assumption of homogeneity across individual Giorgi regions does not hold. Figure 4 shows what each model says is appropriate "amount" of solar reduction to offset temperature or precipitation changes at different points in the South Asia Giorgi region. These spatially and temporally refined results indicate that aggregation obscures results.

However, one cannot determine whether those more granular differences matter by only using a hierarchical modeling approach (Longino, 1990; Sikka, 2018). It is not obvious whether cooler temperatures or less rain are good or bad for communities in particular regions (Sedova et al., 2020). Moreover, hierarchical modeling can miss important conclusions or factors that would be revealed by interacting directly with communities. As an example, Arctic communities are psychologically impacted by increased winter rainfall and decreased snowfall, leading to increased incidence of depression and suicide [I. Mettiainen, personal communication]. Common overextensions of hierarchical modeling include (A) physical climate is a good proxy for impacts like food and water security, or (B) less climate change is better. As a counterexample, interactional modeling has revealed that water resource scarcity may be worsened by greenhouse gas emissions mitigation achieved via increased use of biofuels (Hejazi et al., 2015).

How can solar geoengineering begin to include a more interactional approach? Some obvious low-hanging fruit is to involve more communities who can provide needed perspectives. For example, the field has slowly evolved to include collaborations with experts in various impacts-related fields – agricultural impacts (Fan et al., 2021; Proctor, 2021; Xia et al., 2014), air quality and human health (Eastham et al., 2018; Madronich et al., 2018; Nowack et al., 2016), and ecosystem effects (Trisos et al., 2018; Zarnetske et al., 2021). Another useful step could involve scenario design that incorporates reflexive or participatory modeling (Bistline et al., 2021; McLaren, 2018; Willey, 2016). Involving a variety of stakeholders in the design stage of modeling scenarios, as has been called for in integrated assessment modeling for carbon dioxide removal (Low & Schäfer, 2020; Salter et al., 2010) and solar geoengineering (McLaren, 2018), could address issues of equity in outcomes, in line with the FCE arguments we have presented here. This has the added advantage of increasing transparency of the often poorly documented values and assumptions that go into modeling (Bistline et al., 2021).

Finally, Kravitz et al. found that for temperature, many regions "benefit" from a substantial amount of geoengineering, but for precipitation, at least one region is made worse off by any amount. Presenting either conclusion in isolation would have provided a misleading picture. Instead, Kravitz et al. inadvertently chose an approach more aligned with feminist science by presenting a plurality of outcomes rather than generalizing for the reader and obscuring potentially important conclusions.

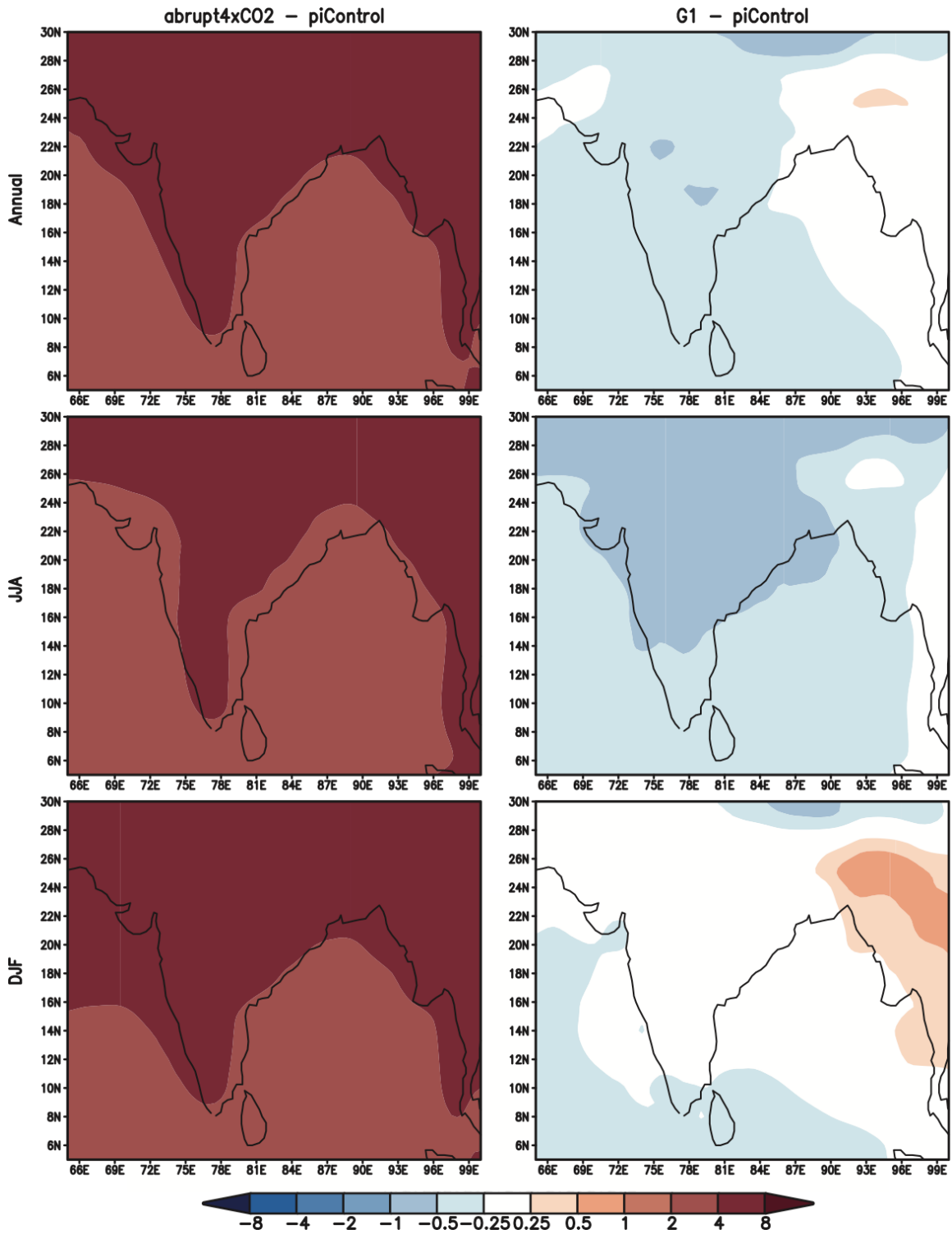

**Figure 2.** Temperature (°C) change in the high $CO_2$ world (left) and a world with high $CO_2$ and solar reduction (right) in the South Asia Giorgi region. Top row shows annual mean changes over a 40-year average of simulation, middle row shows June-July-August average change, and bottom row shows December-January-February average change.

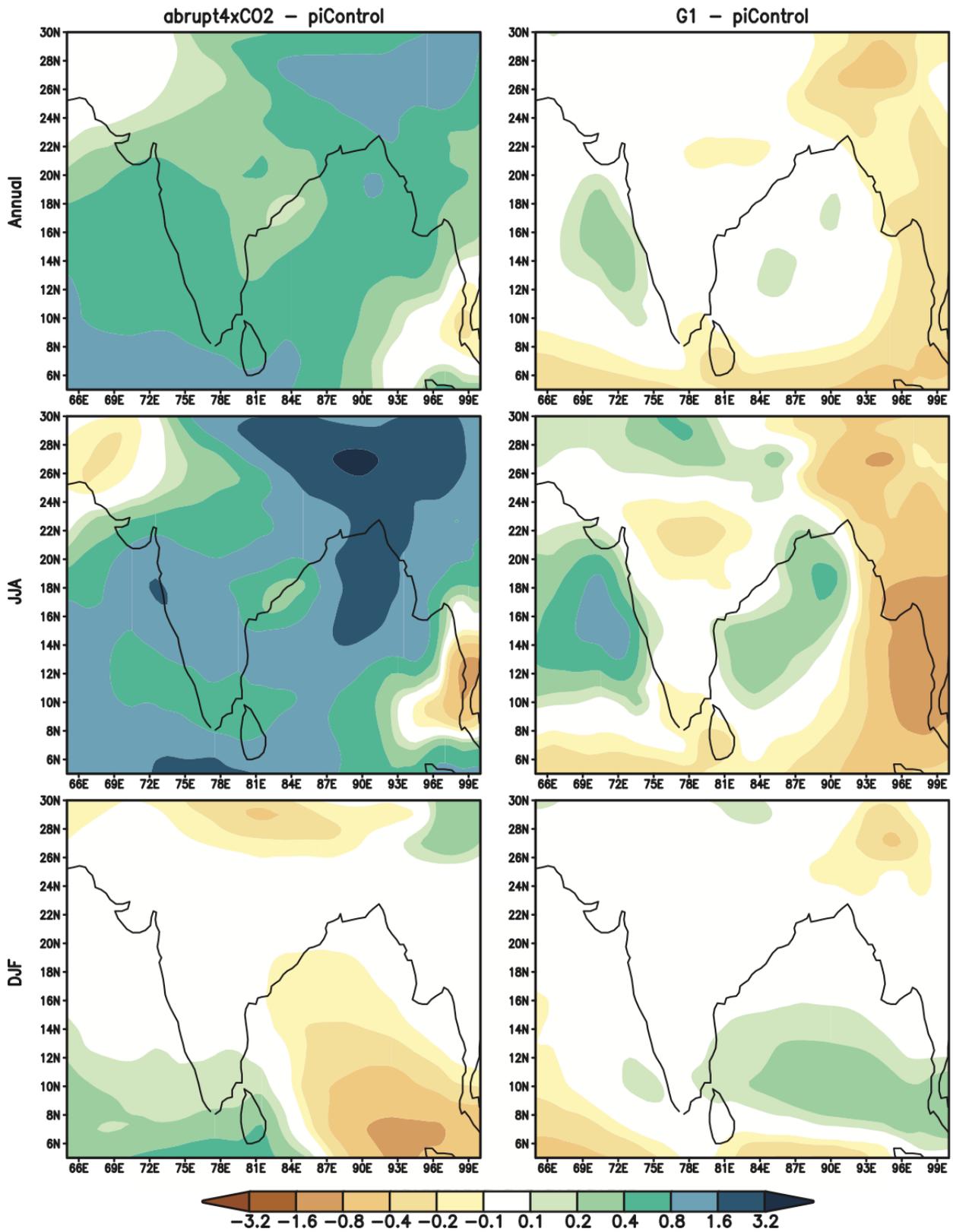

**Figure 3.** As in Figure 2 but for precipitation change (mm/day).

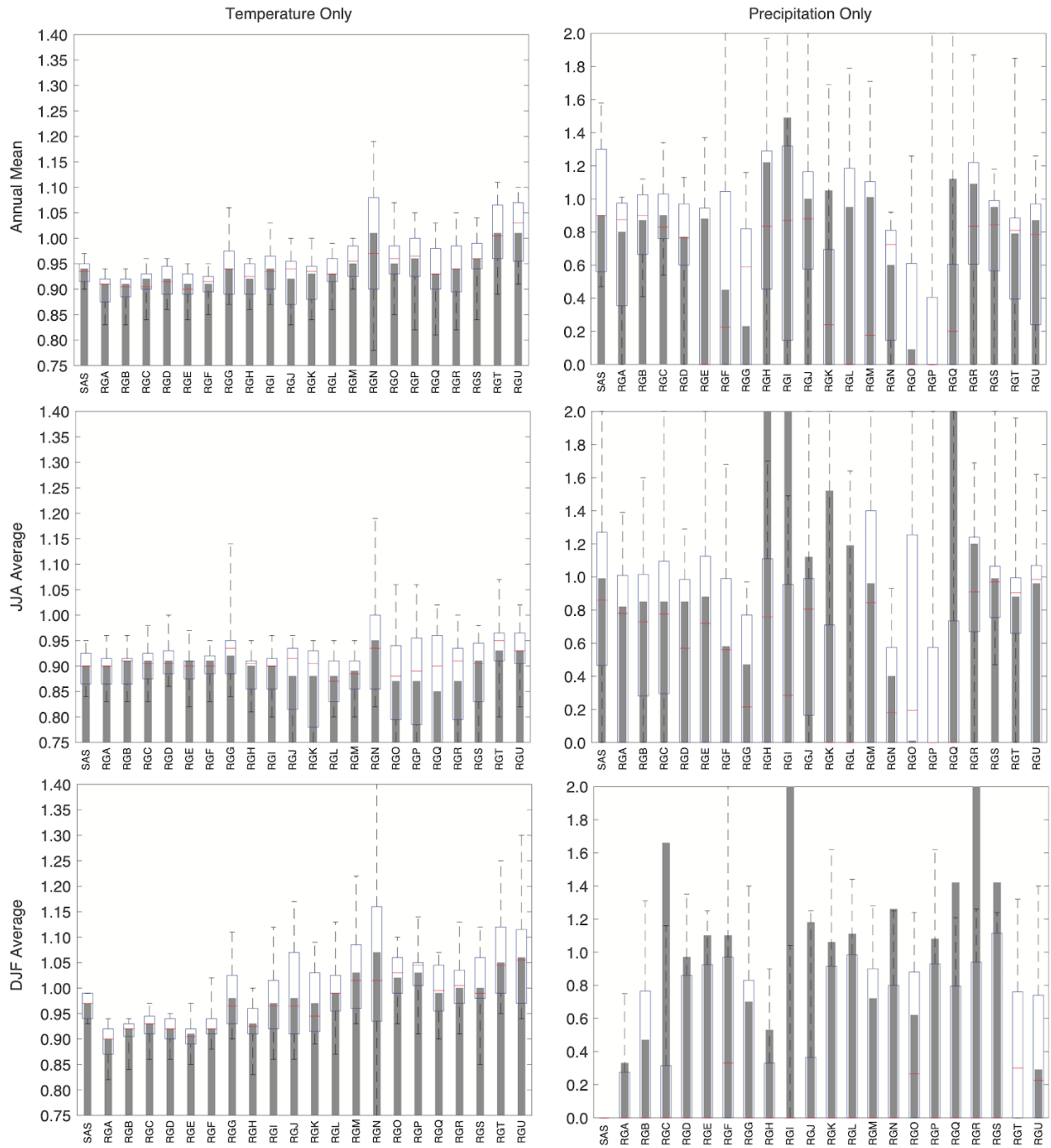

**Figure 4.** The "amount" of geoengineering (solar reduction) required to offset temperature change (left) or precipitation change (right) in South Asia (SAS) or different locations in the South Asia domain (RG#). Amount (y-axis) is defined as a linear scaling where 0 is no solar reduction, and 1 is the amount of solar reduction that will offset global mean temperature change in the annual mean (top), June-July-August mean (middle), or December-January-February mean (bottom). Box and whiskers show the interquartile and range of the 12-model spread, red lines show the median model, and grey bars show the model average. All panels are reprinted from (Kravitz et al., 2014) under a CC BY 3.0 license.

### 3.2.2  What criteria do Kravitz et al. use to justify acceptance of evidence?

Solar geoengineering has not been deployed in the real world, and the use of natural analogues (like volcanoes) has limits, so most results about solar geoengineering are from climate models (Kravitz & MacMartin, 2020).  Even though models are inevitably wrong in some capacity, in the absence of other sources, the evidence provided by models is sometimes accepted as an approximation of what might happen in real world deployments of solar geoengineering (Wieding et al., 2020), even in the highly idealized G1 scenario (Flegal & Gupta, 2018).

A multi-model ensemble average tends to be more accurate than individual model results because various forms of random noise/error get averaged out (Knutti et al., 2010).  Similarly, time averaging reduces noise from interannual variability.  Also, averaging reduces the dimensionality of the results, which makes displaying the results tractable.  Nevertheless, averaging also obscures results. Figure 5 shows that the model average of (time-averaged) temperature or precipitation often has a different sign of response from the minimum or maximum over all models.  Kravitz et al. to a large degree focus on robustness and multi-model agreement, but a feminist science approach would argue that disagreement or outliers could be just as important, especially since there is no strong justification to believe model consensus over individual model results for the simulations they study.

Based on a history of volcanic eruption simulations before there were sophisticated methods of representing aerosol microphysics (Handler, 1989), it is often assumed that solar dimming is a useful proxy for the climate effects of stratospheric sulfate aerosols.  Because solar dimming inaccurately represents the pattern of insolation reduction and stratospheric heating from stratospheric sulfate aerosol geoengineering, there are substantial differences in surface climate response between the two representations, which could lead to erroneous conclusions when evaluating downstream impacts (Visioni et al., 2021).

Comparing simulations of solar geoengineering with the baseline highlights imperfections in how well geoengineering compensates for climate change.  Many have argued that a more fair comparison (and one truer to the intended purpose) is to compare solar geoengineering against a world with high $CO_2$ alone (Govindasamy & Caldeira, 2000; Rasch et al., 2008).  Many GeoMIP studies show both to present a more holistic picture of the results (Kravitz et al., 2013).  Values near zero can be represented as white to obscure small changes, or one can choose color scales that have warm colors for any positive value and cool colors for any negative value.  The latter choice emphasizes differences between geoengineering and the preindustrial baseline, whereas the former downplays differences.  Any attempts to quantify results will require a choice that has advantages and disadvantages, but FCE argues that these choices reflect values and interests that should be interrogated (Sikka, 2018).

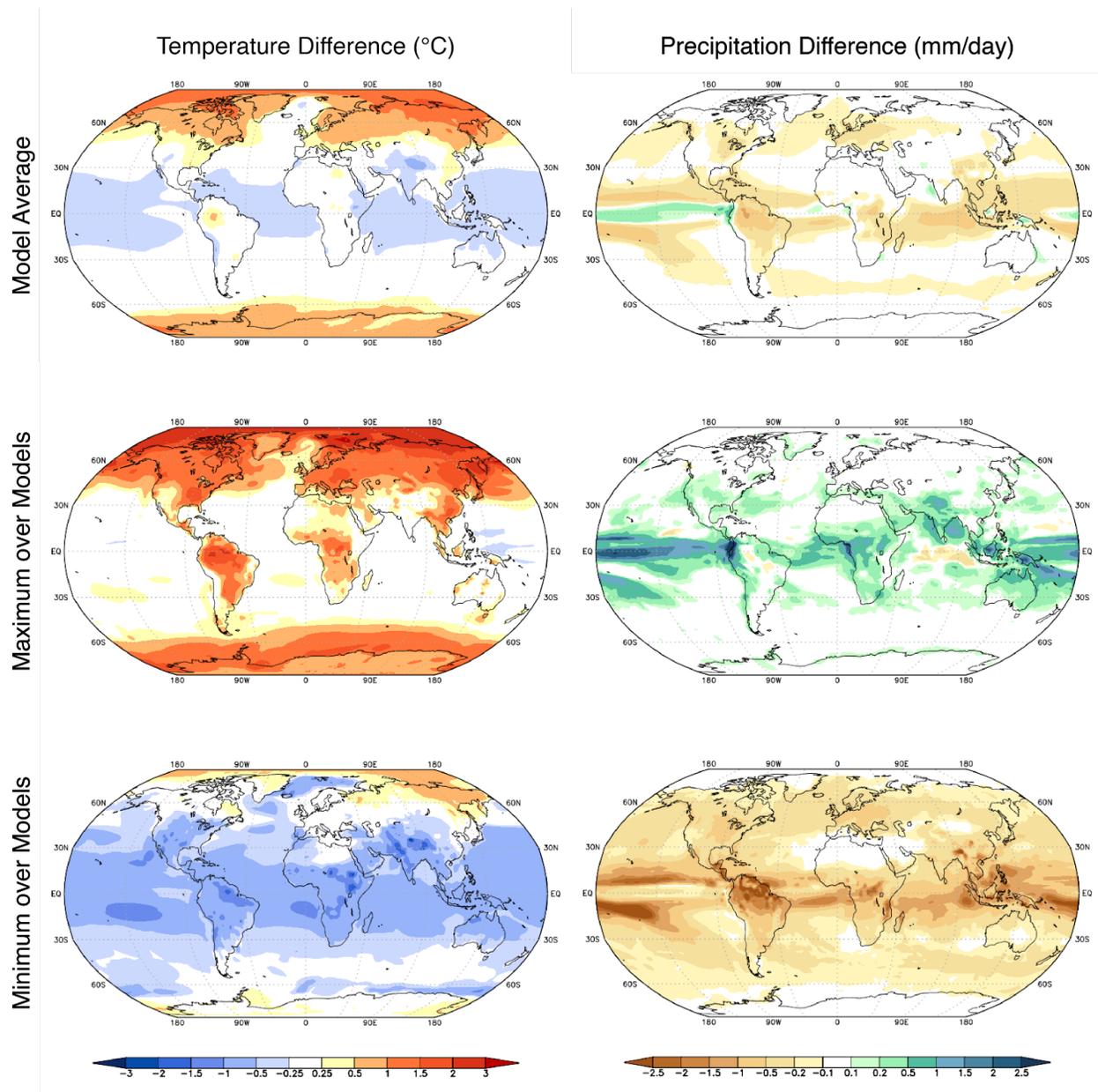

**Figure 5.** Change in temperature (left; °C) and precipitation (right; mm/day) due to solar dimming (GeoMIP experiment G1), as compared to a preindustrial control simulation. Top row shows the 12-model average of the models analyzed by Kravitz et al. (2014). Middle and bottom rows show the maximum and minimum, respectively, over those 12 models on a grid cell basis. Figure combines analyses shown by Kravitz et al. (2013) and Kravitz & MacMartin (2020).

With these issues in mind, we argue against universal standards of analysis, evaluation of uncertainty, or reporting, as any such standards would inadvertently introduce bias into the emergent conclusions from the field. Instead, in line with feminist approaches, we encourage scientists to adopt practices in which consciously identifying choices and understanding impacts is commonplace (Cancian, 1992). We provide some sample questions that scientists could ask themselves when producing a study:

- Am I averaging my results? Would I obtain different conclusions if I averaged a different way?
- Am I reporting results for some regions or time periods in favor of others? Would I obtain different conclusions if I made different choices?
- How am I choosing to plot my results? Would a reader's conclusions change if I plotted the results in a different way?
- On what sources of evidence am I relying? Are there other sources of evidence? Would I obtain different conclusions if I used different sources of evidence? How am I reporting confidence in my results?
- What results am I choosing to report? Would a different scientist, perhaps from a different location or demographic, report different results, or would they report the same results differently?

### 3.2.3 Are there alternate explanations for phenomena?

In simulations that use solar reduction to counteract global warming from $CO_2$, the tropics are "overcooled," and the poles are "undercooled" (Govindasamy & Caldeira, 2000; Kravitz et al., 2013) (also see Figure 5). Numerous studies have attributed this feature to the latitude distribution of forcing: $CO_2$ forcing is ubiquitous, whereas solar reduction has a greater effect in the tropics than the poles. While simple and plausible as an explanation, such statements were effectively unquestioned for a long time. It was recently discovered that meridional energy transport plays an important role in residual polar warming, as well as effects on lapse rate, namely that $CO_2$ tends to enhance high latitude surface warming more than solar irradiance changes (Henry & Merlis, 2020).

Regardless of the mechanism, overcooling/undercooling appears to be a robust result of climate model simulations of solar reduction (Kravitz et al., 2013, 2020). Simulations of equatorial injection of $SO_2$, one of the more commonly simulated representations of stratospheric sulfate aerosol geoengineering, also show overcooling of the tropics and undercooling of the poles (Kravitz et al., 2019). However, simulations involving multiple $SO_2$ injection locations, which are designed to obtain more latitudinally even cooling (Kravitz et al., 2017), are increasingly replacing equatorial injection simulations. A natural question is, even though the wrong explanation persisted throughout the literature for so long, does that matter? Solar dimming poorly represents the climate effects of stratospheric sulfate aerosol geoengineering (Visioni et al., 2021), so correcting erroneous mechanistic understanding of a robust result from a simulation with limited applicability seems moot.

Nevertheless, while this particular error may be of little practical import, failure to consider alternative hypotheses can have important consequences, potentially leading to entire lines of erroneous inquiry. Persistence of incorrect hypotheses can last for a considerable amount of time, which poses serious problems for time-sensitive decisions like how to address climate change and could result in critical decisions being made on false premises. In some cases, this persistence can have important and harmful consequences; as an egregious example, there are numerous instances of declarations that women and minorities are worse at STEMM to justify why they are underrepresented in STEMM fields (Committee on Increasing the Number of Women in Science, Technology, Engineering, Mathematics, and Medicine (STEMM) et al., 2020; Pawley & Hoegh, 2011).

Settling on simple explanations for phenomena is a common form of cognitive bias and was one of the motivations for establishing methods of avoiding these biases, like Analysis of Competing Hypotheses (ACH) (Heuer & Pherson, 2011). While the consequences for settling on this incorrect mechanisms may be low, because of the urgency of climate change, as well as the increasing amount of discussion around solar geoengineering, low consequences cannot be expected in all cases. It may be prudent to attempt to identify and articulate multiple possible theories for explanations of phenomena, even if the eventual answer is clearly in favor of a single hypothesis. A core feminist science value of recognizing heterogeneity represents an accurate reflection of the diversity of hard-to-interpret data and rich uncertainty. Maintaining this as a practice improves transparency in the way science is conducted (Sikka, 2018).

### 3.2.4 What are the values that underpin the science in Kravitz et al.?

There are several implicit or explicit values that guide the analyses of Kravitz et al. Our purpose is not to assess how well those values serve the purposes of Kravitz et al. or how to do better, but instead to name the values and lay bare their implications, which FCE calls for. In doing so, we illustrate where the assumptions or chosen methodologies could have led to knowledge gaps. Kravitz et al. incorporated (and sometimes simultaneously missed opportunities to incorporate) the following values in their analyses:

- Choice of variables (mean temperature and precipitation), which emphasizes regions for which those variables are relevant and deprioritizes others
- Standardization, which improves confidence because it enables multi-model intercomparison but necessitates more idealized scenarios
- Participation, in which all available models were included, with varying degrees of accuracy
- Precedent, focusing on Giorgi regions because past geoengineering studies used them
- Quantification, which allows for more precise answers but also introduces assumptions and caveats that may affect accuracy of the results
- Consensus, emphasizing model agreement and spending less effort diagnosing why models may disagree
- Diversity of model response, which lets readers apply their own contexts rather than implicitly making decisions for them as to what is important
- Transparency, in which the findings are reproducible and assumptions embedded in the study are explicitly described

Several of the values stated above may appear to be incompatible or mutually exclusive, such as consensus and diversity of model response. This is not a flaw of inconsistency, but rather an example of how applying principles of FCE (even if unknowingly at the time) can produce richer conclusions.

We provide more details and description as to the implications of these values.

*Choice of variables.* It would be impossible for a single study to evaluate all variables relevant to the impacts of climate change and geoengineering, so this study limited its analyses to changes in temperature and precipitation. These two variables are important for characterizing a breadth of climate impacts (IPCC, 2014). Although precipitation is not a perfect proxy for moisture availability, it sufficiently represents hydrological cycle changes under both climate change and solar geoengineering (Cheng et al., 2019). Moreover, solar reduction and stratospheric sulfate aerosol geoengineering cannot simultaneously compensate changes in both temperature and precipitation under solar reduction or stratospheric sulfate aerosol geoengineering (Niemeier et al., 2013; Tilmes et al., 2013), so this choice is illustrative of trade-offs. Nevertheless, choosing temperature and precipitation de-prioritizes regions where other variables might have greater relevance. For example, snow cover in high latitude regions or changes in free tropospheric wind shear (the latter of which is relevant for hurricane formation and intensification) may have tangible impacts on vulnerable populations that are not captured by temperature and precipitation. Importantly, while temperature often serves as a useful proxy for a wide variety of climate impacts under climate change (IPCC, 2014), that relationship no longer holds under scenarios with substantial solar geoengineering, as one can suppress temperature change but still have side effects from the combined forcings of greenhouse gases and solar geoengineering (Irvine et al., 2016).

*Standardization.* This study was conducted under the auspices of GeoMIP, meaning that all participating models conducted the same idealized solar reduction scenario. As such, researchers can evaluate places where the models agree (building confidence in those conclusions) or disagree (highlighting areas for further research). Nevertheless, standardization necessitates compromise in scenario design: the scenarios are often idealized (and in the case of G1 idealized to the point that its relevance to stratospheric aerosol geoengineering is questionable) (Visioni et al., 2021) to encourage broad participation and to aid in analysis of the results. Less coordinated efforts have more freedom to explore alternate scenarios for specific purposes but without the confidence gained through a multi-model intercomparison.

*Participation.* Kravitz et al. used an experiment in which (at the time) 12 models participated. All models were weighted equally, without an attempt to determine whether there should be greater or less confidence in any one model's results. Moreover, while there are efforts to weight models or eliminate poor-performing models in studies of climate change (Tebaldi et al., 2011), because there are no real-world observations of geoengineering, there is less basis on which GeoMIP studies could perform a similar evidence-based weighting. As such, participating models may vary in the accuracy of their temperature or precipitation responses to forcing.

*Precedent.* Much of the analysis by Kravitz et al. focuses on Giorgi regions for ease of analysis and to improve robustness of the results, as analyzing each model grid point separately would be cumbersome, and spatially aggregated results are often more trustworthy than results from individual grid boxes. Nevertheless, there are numerous choices for spatial aggregation that could have been pursued, including geopolitical boundaries, Köppen-Geiger climate regions, or even bands of latitudes on different continents. In large part, the Giorgi regions were chosen for this study because they were used by Ricke et al. (2010) in a previous geoengineering study. The Giorgi regions have been used in numerous other geoengineering studies, although it would be difficult to argue that their use has been disproportionate compared to other climate research fields.

*Quantification.* Kravitz et al. focused on providing quantitative results regarding regional disparities, including a weighting between temperature and precipitation changes. This required several assumptions and methodology choices that affected the results in a variety of ways that have not been thoroughly studied.

- The scenario G1 was compared to a scenario with a quadrupled $CO_2$ concentration, resulting in a high signal-to-noise ratio but running the risk of exciting climate system nonlinearities, which may result in regional changes that are not representative of moderate climate change or solar geoengineering deployments.
- The "amount" of geoengineering to meet particular outcomes was obtained through linear scaling of the climate effects. This invariably introduced error into the results, although the error is likely greater for extreme scenarios rather than scenarios near the preindustrial baseline.
- To incorporate temperature and precipitation into the same metric, Kravitz et al. normalized changes by the standard deviation of interannual variability, as has been done before (Ricke et al., 2010). This skews the metric toward temperature changes, as the interannual standard deviation of temperature is substantially smaller than that of precipitation.
- They included a quadratic "dis-utility" function (Equation 1), which disproportionately penalizes large departures from the baseline of comparison. This may be appropriate (Nordhaus, 2017) but also places more emphasis on the choice of baseline than a linear metric. Also, it contains the assumption of symmetry, in that "undercooling" by X degrees (as compared to the baseline) is just as damaging as "overcooling" by X degrees.
- They averaged over 22 Giorgi regions, which certainly impacted the results (Figures 2-4).

*Consensus.* Many GeoMIP studies tend to focus on model agreement, evaluating the ensemble average or narrowness of the range of model responses (Kravitz et al., 2013). This reflects the core value of consensus within GeoMIP: evaluating where models agree and where they disagree (Kravitz et al., 2011). Kravitz et al. also emphasized consensus in the form of a Pareto-improving criterion (Moreno-Cruz et al., 2012): how much can the amount of solar geoengineering be increased before any one of the 22 regions is harmed?

*Diversity of model response.* Although some results were obscured by averaging, the results included presentation of model diversity (e.g., Figures 1 and 4). Other disciplines have similarly heterogeneous approaches to reporting results. For example, health impacts of climate change often incorporate climate model information on outcomes but rarely include local knowledge or values-driven community data, making it impossible to accurately prioritize information, provide meaning, and decide on courses of action (Donatuto et al., 2020). Allowing for a diversity of model responses lets readers apply their own contexts instead of implicitly deciding for readers by aggregating the results.

*Transparency.* The value of transparency is simultaneously demonstrated and not demonstrated by Kravitz et al. Descriptions of the methodology and findings are transparent and reproducible, and the analysis methods have been used by other studies (Irvine et al., 2019). However, many other value-laden assumptions are embedded in the study but not explicitly described, meaning they cannot be easily examined for inadvertent effects on the results or conclusions. There is potential for some assumptions to have serious ethical implications (Beck & Krueger, 2016). This issue is widespread and certainly not limited to solar geoengineering (Bistline et al., 2021).

### 3.2.5  Whose interests do the phenomena Kravitz et al. decided to study serve?

Kravitz et al. aimed to resolve uncertainty (see Section 3.2.7), but uncertainty for whom? Scientists want to improve the knowledge base, whereas policy makers want to reduce chances of making political mistakes; miscommunications in the policy making process result from these different conceptions (Enserink et al., 2013). These difficulties are exacerbated when scientists advocate for policy to influence decision making. By evaluating some of the uncertainties addressed by Kravitz et al., we can better understand what interests they are attempting to serve and how successful they were:

- *Climate impacts.*  What are the temperature and precipitation impacts, on a regional basis, of different amounts of solar geoengineering?
- *Model spread.*  Do different models agree on the climate impacts of different amounts of solar geoengineering?  Do they disagree?  Where?
- *Environmental equity.*  Would solar geoengineering result in winners and losers?
- *Environmental justice.*  Could solar geoengineering alleviate suffering from climate change?  For whom?  With what consequences?

Climate models are clearly adept at addressing the first two. For the latter two, climate models may be important, but they are missing fundamental pieces that would allow them to address equity or justice, for example impacts assessment, integrated human-societal modeling, and rigorous policy analysis. Inadvertent overreach could be circumvented by consciously evaluating the scope of the study: what questions do we want to answer, and do we have the right tool?

There are three prevalent "cognitive frames" in which climate scenarios sit (Haikola et al., 2019): discussions of possible future scenarios, political prescriptions that attempt to force particular future scenarios/policies, and distortions of science. In which frame(s) are Kravitz et al. operating when they conducted their study, and in which frame(s) are their results being used? This is not simply a thought exercise: by conducting simulations, scientists are telling policy-shaping stories with those models (Beck, 2018); such narrative structures (constrained by the rules encoded into the models) are unavoidable components of modeling (Morgan, 2001). Because of the ubiquity of climate modeling in solar geoengineering, modeling is responsible for how solar geoengineering is understood (Wiertz, 2016).

Kravitz et al. never explicitly decided upon sociopolitical aims or policy questions for their study, and doing so might have shaped the way in which they conducted their analyses or the results they chose to show. It is clear that others have used physical science studies as evidence for arguments about environmental justice issues in solar geoengineering, sometimes to take positions against solar geoengineering deployment (McKinnon, 2020; Surprise, 2020) or research (Stephens & Surprise, 2020). By implicitly framing their study about winners/losers, Kravitz et al. inadvertently entered into a debate about environmental justice issues in solar geoengineering and, in many senses, helped shape that debate through the futures and outcomes they chose to model. We are not arguing that scientists can avoid policy discussions, and indeed science can be an important part of the evidence basis around which policy is made. Rather, we point out that when conducting studies, it is important to critically evaluate what stories the study is telling. Some key questions Kravitz et al. could have asked include:

- What story is this study telling about winners and losers?
- Are winners and losers from solar geoengineering inevitable?
- Is this issue a potential showstopper for solar geoengineering?
- When describing the results, how much attention should be paid to the winners, and how much to the losers?
- Is there actually a binary between winners and losers?

This list is far from comprehensive, but it is aimed at recognizing the power of narrative: models reproduce and reinforce certain discourses and silence others (Ellenbeck & Lilliestam, 2019). By focusing on what a model simulation represents, not just its output, assumptions can be laid bare for scrutiny.

### 3.2.6 How do Kravitz et al. describe science?

Choice of language, which results to emphasize or de-emphasize, and interpretation all have some amount of subjectivity that will influence any study's messages. (This aspect of value-laden language is one of the founding motivations behind FCE.) For example, primarily discussing average results obscures marginalized groups in favor of an "average experience" that may not actually be realized by anyone. Instead, focusing on disaggregation (Figures 2-4), the local, and the marginalized, as is encouraged by feminist practice, provides a richer picture of the outcomes (Harding, 1991).

Relatedly, the discussion of regional disparities or regional effects also reflects certain choices over others. Alternatives to the Giorgi regions could have included geopolitical boundaries, socioeconomic differences, or cultural/demographic separations. These alternatives could have different approaches to issues of power, for example colonial structures, inequality, and marginalization. Most climate scenarios do not incorporate well-known gender, race, class, and other sociodemographic differences (Kaijser & Kronsell, 2014), and focusing on these instead of more technically oriented boundaries might have produced different assessments of disparities that result from solar geoengineering. Doing so could also improve the degree to which their study evaluates environmental justice.

Kravitz et al. primarily focus on quantitative results, which narrows conclusions, as qualitative results can provide information about material and experiential effects and impacts. Feminist science argues that favoring quantitative results is an effect of gender bias: qualitative methods are often gendered or coded feminine and thus undervalued (Lawson, 1995). However, qualitative descriptions can be overused: for example, at one point Kravitz et al. discuss "small changes" in precipitation, which invariably minimizes certain perspectives: even a "small" change in precipitation might be crucial for subsistence farmers.

### 3.2.7 For what ends are Kravitz et al. practicing science?

No scientific study is purely motivated by curiosity and knowledge production. Scientists have many choices about how they could spend their limited time, and presentation of a completed study implicitly contains choices about what to include (or not).

A winners/losers framing implies assessments of distributional justice and alleviation of misery from climate change. Success in this aim could look like identifying the potential for geopolitical strife (who would be affected and how?) or providing specific, actionable outcomes. The regions being disadvantaged by geoengineering under the chosen metrics in the study differed amongst models, and because the regions were not based on geopolitical, socioeconomic, or cultural/demographic boundaries, there is little potential for the study to result in anything actionable (which, given the idealized nature of the study, might be seen as a benefit so as to avoid misinterpretation of the results as being policy-relevant).

Kravitz et al. built off earlier work (Ricke et al., 2010) to refine uncertainty bounds around regional differences in outcomes from solar geoengineering and to build confidence in their conclusions. Keeping in mind the caveats related to limited sources of evidence, model intercomparisons are effective in building confidence in conclusions by quantifying structural uncertainty in climate model simulations of solar geoengineering.

## 4  A Path Toward More Inclusive Solar Geoengineering Research

Through this feminist science framework and its application to a retrospective analysis of Kravitz et al. (2014), we have identified numerous ways in which their findings could be reanalyzed or reinterpreted. We have also identified underlying assumptions which, if challenged, could lead to different conclusions. Our purpose here is not to fault the study or its authors, as the results have invariably improved our understanding of some of the climatic effects of solar geoengineering. Instead, we illustrate the application of practices that could improve the body of knowledge produced by studies and improve transparency of the way studies are conducted using feminist science. By making a study's values explicit, scientists can assess gaps and opportunities to consider a wider range of values and perspectives. Pluralism, as is encouraged by feminist practice, leads to a more holistic set of scientific conclusions.

Important strides to include multiple perspectives are being pursued. The Carnegie Climate Governance Initiative (C2G) has supported a proposed United Nations resolution on solar geoengineering governance; the International Red Cross/Red Crescent has been involved in rural international public engagement around solar geoengineering for years (Suarez & van Aalst, 2017); and there has been public engagement around solar geoengineering in the Arctic (Buck et al., 2016). In addition, the Developing Country Impacts Modeling Analysis for SRM (DECIMALS) fund aims at building developing country research capacity in solar geoengineering so that these countries will have their own expertise to address issues specific to their perspectives and will be empowered at international discussions (Rahman et al., 2018);

several DECIMALS studies focused on country-level and regional climate and climate extremes have already been produced, with more on the way (e.g., Da-Allada et al., 2020; Pinto et al., 2020). While not sufficient, these steps will result in substantially more plurality in solar geoengineering research and tangible benefits for participatory governance.

The discussions here are hopefully a first step toward a more thorough and rigorous application of this feminist science-based framework to solar geoengineering. The questions raised here are a good start but incomplete. Moreover, not all of the questions or suggestions may be applicable to every study. Further retrospective analyses, as well as applications to future studies, could help refine these lists into a set of best scientific practices. Ultimately, the purpose is to consciously evaluate the values being encapsulated by a study to understand how they (or the absence of other values) might shape conclusions. Regardless of the outcome of this evaluation process (for example, the implicit values might have a neutral or positive effect on the study), we argue that the answer is important to know.

**Author Contributions**

Both authors contributed to designing the study and writing the paper.

**Acknowledgments**

We thank Miranda Böttcher for invaluable comments on the manuscript.

**Funding**

Support for B.K. was provided in part by the National Science Foundation through agreement CBET-1931641, the Indiana University Environmental Resilience Institute, and the *Prepared for Environmental Change* Grand Challenge initiative. The Pacific Northwest National Laboratory is operated for the US Department of Energy by Battelle Memorial Institute under contract DE-AC05-76RL01830.

**Competing Interests**

The authors have declared that no competing interests exist.

**Data Accessibility Statement**

Data are available via the Earth System Grid Federation; see instructions at http://www.geomip.org.

# References

Beck, M. (2018). Telling stories with models and making policy with stories: An exploration. *Climate Policy*, *18*(7), 928–941. https://doi.org/10.1080/14693062.2017.1404439

Beck, M., & Krueger, T. (2016). The epistemic, ethical, and political dimensions of uncertainty in integrated assessment modeling: The epistemic, ethical, and political dimensions of uncertainty in integrated assessment modeling. *Wiley Interdisciplinary Reviews: Climate Change*, *7*(5), 627–645. https://doi.org/10.1002/wcc.415

Beck, S., & Mahony, M. (2018). The politics of anticipation: The IPCC and the negative emissions technologies experience. *Global Sustainability*, *1*, e8. https://doi.org/10.1017/sus.2018.7

Biermann, F., & Möller, I. (2019). Rich man's solution? Climate engineering discourses and the marginalization of the Global South. *International Environmental Agreements: Politics, Law and Economics*, *19*(2), 151–167. https://doi.org/10.1007/s10784-019-09431-0

Bijker, W. (2017). Constructing Worlds: Reflections on Science, Technology and Democracy (and a Plea for Bold Modesty). *Engaging Science, Technology, and Society*, *3*, 315. https://doi.org/10.17351/ests2017.170

Bistline, J., Budolfson, M., & Francis, B. (2021). Deepening transparency about value-laden assumptions in energy and environmental modelling: Improving best practices for both modellers and non-modellers. *Climate Policy*, *21*(1), 1–15. https://doi.org/10.1080/14693062.2020.1781048

Buck, H. J., Gammon, A. R., & Preston, C. J. (2014). Gender and Geoengineering. *Hypatia*, *29*(3), 651–669. https://doi.org/10.1111/hypa.12083

Buck, H. J., Mettiainen, I., MacMartin, D., & Ricke, K. (2016). *Deliberating Albedo Modification in Finnish Lapland: Integrating Geoengineering Research With Community-Specific Insights*. American Geophysical Union Fall Meeting 2016. https://ui.adsabs.harvard.edu/abs/2016AGUFMGC33B1236B/abstract

Calvin, K., & Bond-Lamberty, B. (2018). Integrated human-earth system modeling—State of the science and future directions. *Environmental Research Letters*, *13*(6), 063006. https://doi.org/10.1088/1748-9326/aac642

Campbell, M. C., Vogel, M., & Williams, J. (2015). HISTORICAL CONTINGENCIES AND THE EVOLVING IMPORTANCE OF RACE, VIOLENT CRIME, AND REGION IN EXPLAINING MASS INCARCERATION IN THE UNITED STATES: EXPLAINING INCARCERATION IN THE UNITED STATES. *Criminology*, *53*(2), 180–203. https://doi.org/10.1111/1745-9125.12065

Cancian, F. M. (1992). FEMINIST SCIENCE: Methodologies that Challenge Inequality. *Gender & Society*, *6*(4), 623–642. https://doi.org/10.1177/089124392006004006

Cheng, W., MacMartin, D. G., Dagon, K., Kravitz, B., Tilmes, S., Richter, J. H., Mills, M. J., & Simpson, I. R. (2019). Soil Moisture and Other Hydrological Changes in a Stratospheric Aerosol Geoengineering Large Ensemble. *Journal of Geophysical Research: Atmospheres*, *124*(23), 12773–12793. https://doi.org/10.1029/2018JD030237

Committee on Developing a Research Agenda and Research Governance Approaches for Climate Intervention Strategies that Reflect Sunlight to Cool Earth, Board on Atmospheric Sciences and Climate, Committee on Science, Technology, and Law, Division on Earth and Life Studies, Policy and Global Affairs, & National Academies of Sciences, Engineering, and Medicine. (2021). *Reflecting Sunlight: Recommendations for Solar Geoengineering Research and Research Governance* (p. 25762). National Academies Press. https://doi.org/10.17226/25762

Committee on Increasing the Number of Women in Science, Technology, Engineering, Mathematics, and Medicine (STEMM), Committee on Women in Science, Engineering, and Medicine, Policy and Global Affairs, & National Academies of Sciences, Engineering, and Medicine. (2020). *Promising Practices for Addressing the Underrepresentation of Women in Science, Engineering, and Medicine: Opening Doors* (R. Colwell, A. Bear, & A. Helman, Eds.; p. 25585). National Academies Press. https://doi.org/10.17226/25585

Crutzen, P. J. (2006). Albedo enhancement by stratospheric sulfur injections: A contribution to resolve a policy dilemma? *Climatic Change*, *77*, 211–220. https://doi.org/10.1007/s10584-006-9101-y

Da-Allada, C. Y., Baloïtcha, E., Alamou, E. A., Awo, F. M., Bonou, F., Pomalegni, Y., Biao, E. I., Obada, E., Zandagba, J. E., Tilmes, S., & Irvine, P. J. (2020). Changes in West African Summer Monsoon Precipitation Under Stratospheric Aerosol Geoengineering. *Earth's Future*, *8*(7), e2020EF001595. https://doi.org/10.1029/2020EF001595

DeLoughrey, E., Didur, J., & Carrigan, A. (2015). *Global ecologies and the environmental humanities: Postcolonial approaches*. Routledge.

Donatuto, J., Campbell, L., & Trousdale, W. (2020). The "value" of values-driven data in identifying Indigenous health and climate change priorities. *Climatic Change*, *158*(2), 161–180. https://doi.org/10.1007/s10584-019-02596-2

Eastham, S. D., Weisenstein, D. K., Keith, D. W., & Barrett, S. R. H. (2018). Quantifying the impact of sulfate geoengineering on mortality from air quality and UV-B exposure. *Atmospheric Environment*, *187*, 424–434. https://doi.org/10.1016/j.atmosenv.2018.05.047

Ellenbeck, S., & Lilliestam, J. (2019). How modelers construct energy costs: Discursive elements in Energy System and Integrated Assessment Models. *Energy Research & Social Science*, *47*, 69–77. https://doi.org/10.1016/j.erss.2018.08.021

Enserink, B., Kwakkel, J. H., & Veenman, S. (2013). Coping with uncertainty in climate policy making: (Mis)understanding scenario studies. *Futures*, *53*, 1–12. https://doi.org/10.1016/j.futures.2013.09.006

Fan, Y., Tjiputra, J., Muri, H., Lombardozzi, D., Park, C.-E., Wu, S., & Keith, D. (2021). Solar geoengineering can alleviate climate change pressures on crop yields. *Nature Food*, *2*(5), 373–381. https://doi.org/10.1038/s43016-021-00278-w

Flegal, J. A., & Gupta, A. (2018). Evoking equity as a rationale for solar geoengineering research? Scrutinizing emerging expert visions of equity. *International Environmental Agreements: Politics, Law and Economics*, *18*(1), 45–61. https://doi.org/10.1007/s10784-017-9377-6

Gilbert, S., & Rader, K. (1998). How does gender matter? Revisiting women, feminism and developmental biology. *Feminism in Twentieth-Century Science, Technology and Medicine*. Science, medicine, and technology in the 20th century: What difference has feminism made?

Giorgi, F., & Francisco, R. (2000). Evaluating uncertainties in the prediction of regional climate change. *Geophysical Research Letters*, *27*(9), 1295–1298. https://doi.org/10.1029/1999GL011016

Govindasamy, B., & Caldeira, K. (2000). Geoengineering Earth's radiation balance to mitigate $CO_2$-induced climate change. *Geophys. Res. Lett.*, *27*, 2141–2144. https://doi.org/10.1029/1999GL006086

Gupta, A., & Möller, I. (2019). De facto governance: How authoritative assessments construct climate engineering as an object of governance. *Environmental Politics*, *28*(3), 480–501. https://doi.org/10.1080/09644016.2018.1452373

Haikola, S., Hansson, A., & Fridahl, M. (2019). Map-makers and navigators of politicised terrain: Expert understandings of epistemological uncertainty in integrated assessment modelling of bioenergy with carbon capture and storage. *Futures*, *114*, 102472. https://doi.org/10.1016/j.futures.2019.102472

Handler, P. (1989). The effect of volcanic aerosols on global climate. *Journal of Volcanology and Geothermal Research*, *37*(3–4), 233–249. https://doi.org/10.1016/0377-0273(89)90081-4

Harding, S. G. (1991). *Whose science? Whose knowledge? thinking from women's lives*. Cornell University Press.

Hejazi, M. I., Voisin, N., Liu, L., Bramer, L. M., Fortin, D. C., Hathaway, J. E., Huang, M., Kyle, P., Leung, L. R., Li, H.-Y., Liu, Y., Patel, P. L., Pulsipher, T. C., Rice, J. S., Tesfa, T. K., Vernon, C. R., & Zhou, Y. (2015). 21st century United States emissions mitigation could increase water stress more than the climate change it is mitigating. *Proceedings of the National Academy of Sciences*, *112*(34), 10635–10640. https://doi.org/10.1073/pnas.1421675112

Henry, M., & Merlis, T. M. (2020). Forcing Dependence of Atmospheric Lapse Rate Changes Dominates Residual Polar Warming in Solar Radiation Management Climate Scenarios. *Geophysical Research Letters*, *47*(15). https://doi.org/10.1029/2020GL087929

Heuer, R. J., & Pherson, R. H. (2011). *Structured analytic techniques for intelligence analysis*. CQ Press.

Heutel, G., Moreno-Cruz, J., & Shayegh, S. (2016). Climate tipping points and solar geoengineering. *Journal of Economic Behavior & Organization*, *132*, 19–45. https://doi.org/10.1016/j.jebo.2016.07.002

Hoppe, R. (2005). Rethinking the science-policy nexus: From knowledge utilization and science technology studies to types of boundary arrangements. *Poiesis & Praxis*, *3*(3), 199–215. https://doi.org/10.1007/s10202-005-0074-0

IPCC. (2013). *Climate Change 2013: The Physical Science Basis. Contribution of Working Group I to the Fifth Assessment Report of the Intergovernmental Panel on Climate Change*. Cambridge University Press. https://doi.org/10.1017/CBO9781107415324

IPCC. (2014). *Climate Change 2014: Impacts, Adaptation, and Vulnerability. Contribution of Working Group II to the Fifth Assessment Report of the Intergovernmental Panel on Climate Change*. Cambridge University Press. www.ipcc.ch/report/ar5/wg2

Irvine, P., Emanuel, K., He, J., Horowitz, L. W., Vecchi, G., & Keith, D. (2019). Halving warming with idealized solar geoengineering moderates key climate hazards. *Nature Climate Change*, *9*(4), 295–299. https://doi.org/10.1038/s41558-019-0398-8

Irvine, P. J., Kravitz, B., Lawrence, M. G., & Muri, H. (2016). An overview of the Earth system science of solar geoengineering. *WIREs Climate Change*, *7*, 815–833. https://doi.org/10.1002/wcc.423

Kaijser, A., & Kronsell, A. (2014). Climate change through the lens of intersectionality. *Environmental Politics*, *23*(3), 417–433. https://doi.org/10.1080/09644016.2013.835203

Keith, D. W. (2013). *A case for climate engineering*. The MIT Press.

Keith, D. W., & Irvine, P. J. (2016). Solar geoengineering could substantially reduce climate risks-A research hypothesis for the next decade: SOLAR GEOENGINEERING COULD REDUCE RISK. *Earth's Future*, *4*(11), 549–559. https://doi.org/10.1002/2016EF000465

Keller, E. F. (2002). *Making Sense of Life*. Harvard University Press.

Knutti, R., Furrer, R., Tebaldi, C., Cermak, J., & Meehl, G. A. (2010). Challenges in Combining Projections from Multiple Climate Models. *Journal of Climate*, *23*(10), 2739–2758. https://doi.org/10.1175/2009JCLI3361.1

Kravitz, B., Caldeira, K., Boucher, O., Robock, A., Rasch, P. J., Alterskjaer, K., Karam, D. B., Cole, J. N. S., Curry, C. L., Haywood, J. M., Irvine, P. J., Ji, D., Jones, A., Kristjánsson, J. E., Lunt, D. J., Moore, J. C., Niemeier, U., Schmidt, H., Schulz, M., … Yoon, J.-H. (2013). Climate model response from the Geoengineering Model Intercomparison Project (GeoMIP). *Journal of Geophysical Research: Atmospheres*, *118*(15), 8320–8332. https://doi.org/10.1002/jgrd.50646

Kravitz, B., & MacMartin, D. G. (2020). Uncertainty and the basis for confidence in solar geoengineering research. *Nature Reviews Earth & Environment*, *1*(1), 64–75. https://doi.org/10.1038/s43017-019-0004-7

Kravitz, B., MacMartin, D. G., Mills, M. J., Richter, J. H., Tilmes, S., Lamarque, J., Tribbia, J. J., & Vitt, F. (2017). First Simulations of Designing Stratospheric Sulfate Aerosol Geoengineering to Meet Multiple Simultaneous Climate Objectives. *Journal of Geophysical Research: Atmospheres*, *122*(23). https://doi.org/10.1002/2017JD026874

Kravitz, B., MacMartin, D. G., Robock, A., Rasch, P. J., Ricke, K. L., Cole, J. N. S., Curry, C. L., Irvine, P. J., Ji, D., Keith, D. W., Egill Kristjánsson, J., Moore, J. C., Muri, H., Singh, B., Tilmes, S., Watanabe, S., Yang, S., & Yoon, J.-H. (2014). A multi-model assessment of regional climate disparities caused by solar geoengineering. *Environmental Research Letters*, *9*(7), 074013. https://doi.org/10.1088/1748-9326/9/7/074013

Kravitz, B., MacMartin, D. G., Tilmes, S., Richter, J. H., Mills, M. J., Cheng, W., Dagon, K., Glanville, A. S., Lamarque, J., Simpson, I. R., Tribbia, J., & Vitt, F. (2019). Comparing Surface and Stratospheric Impacts of Geoengineering With Different SO $_2$ Injection Strategies. *Journal of Geophysical Research: Atmospheres*, *124*(14), 7900–7918. https://doi.org/10.1029/2019JD030329

Kravitz, B., MacMartin, D. G., Visioni, D., Boucher, O., Cole, J. N. S., Haywood, J., Jones, A., Lurton, T., Nabat, P., Niemeier, U., Robock, A., Séférian, R., & Tilmes, S. (2020). *Comparing different generations of idealized solar geoengineering simulations in the Geoengineering Model Intercomparison Project (GeoMIP)* [Preprint]. Aerosols/Atmospheric Modelling/Stratosphere/Physics (physical properties and processes). https://doi.org/10.5194/acp-2020-732

Kravitz, B., Robock, A., Boucher, O., Schmidt, H., Taylor, K. E., Stenchikov, G., & Schulz, M. (2011). The Geoengineering Model Intercomparison Project (GeoMIP). *Atmospheric Science Letters*, *12*(2), 162–167. https://doi.org/10.1002/asl.316

Kuhn, T. S. (1977). Objectivity, Value Judgment, and Theory Choice. In *Arguing about Science* (pp. 74–86).

Lawson, V. (1995). The Politics of Difference: Examining the Quantitative/Qualitative Dualism in Post-Structuralist Feminist Research. *The Professional Geographer*, *47*(4), 449–457. https://doi.org/10.1111/j.0033-0124.1995.449_1.x

Liboiron, M., Duman, N., Bond, A., Charron, L., Liboiron, F., Ammendolia, J., Hawkins, K., Wells, E., Melvin, J., Dawe, N., & Novacefski, M. (2020). *Regional report on plastic pollution in Newfoundland and Labrador, 1962-2019*. Civic Laboratory for Environmental Action Research. https://civiclaboratory.files.wordpress.com/2020/09/clear-regional-report-on-plastic-pollution-in-nl-1962-2019.pdf

Lloyd, E. A. (2009). *Case of the Female Orgasm: Bias in the Science of Evolution*. Harvard University Press. http://qut.eblib.com.au/patron/FullRecord.aspx?p=3300297

Longino, H. (2019). The Social Dimensions of Scientific Knowledge. In *Stanford Encyclopedia of Philosophy*. Stanford University. https://plato.stanford.edu/archives/sum2019/entries/scientific-knowledge-social/

Longino, H. E. (1987). Can There Be A Feminist Science? *Hypatia*, *2*(3), 51–64. Cambridge Core. https://doi.org/10.1111/j.1527-2001.1987.tb01341.x

Longino, H. E. (1990). *Science as social knowledge: Values and objectivity in scientific inquiry*. Princeton University Press.

Longino, H. E. (1996). Cognitive and Non-Cognitive Values in Science: Rethinking the Dichotomy. In L. H. Nelson & J. Nelson (Eds.), *Feminism, Science, and the Philosophy of Science* (pp. 39–58). Springer Netherlands. https://doi.org/10.1007/978-94-009-1742-2_3

Longino, H. E., & Lennon, K. (1997). Feminist Epistemology as a Local Epistemology. *Proceedings of the Aristotelian Society, Supplementary Volumes*, *71*, 19–54. JSTOR.

Low, S., & Schäfer, S. (2020). Is bio-energy carbon capture and storage (BECCS) feasible? The contested authority of integrated assessment modeling. *Energy Research & Social Science*, *60*, 101326. https://doi.org/10.1016/j.erss.2019.101326

Madronich, S., Tilmes, S., Kravitz, B., MacMartin, D., & Richter, J. (2018). Response of Surface Ultraviolet and Visible Radiation to Stratospheric SO2 Injections. *Atmosphere*, *9*(11), 432. https://doi.org/10.3390/atmos9110432

Martin, E. (1991). The egg and the sperm: How science has constructed a romance based on stereotypical male-female roles. *Signs: Journal of Women in Culture and Society*, *16*(3), 485–501.

McKinnon, C. (2020). The Panglossian politics of the geoclique. *Critical Review of International Social and Political Philosophy*, *23*(5), 584–599. https://doi.org/10.1080/13698230.2020.1694216

McLaren, D. P. (2018). Whose climate and whose ethics? Conceptions of justice in solar geoengineering modelling. *Energy Research & Social Science*, *44*, 209–221. https://doi.org/10.1016/j.erss.2018.05.021

Metoyer, A. B., & Rust, R. (2011). The Egg, Sperm, and Beyond: Gendered Assumptions in Gynecology Textbooks. *Women's Studies*, *40*(2), 177–205. https://doi.org/10.1080/00497878.2011.537986

Moreno-Cruz, J. B., Ricke, K. L., & Keith, D. W. (2012). A simple model to account for regional inequalities in the effectiveness of solar radiation management. *Climatic Change*, *110*(3–4), 649–668. https://doi.org/10.1007/s10584-011-0103-z

Morgan, M. S. (2001). Models, stories and the economic world. *Journal of Economic Methodology*, *8*(3), 361–384. https://doi.org/10.1080/13501780110078972

National Research Council. (2015a). *Climate Intervention: Carbon Dioxide Removal and Reliable Sequestration*. The National Academies Press. https://doi.org/10.17226/18805

National Research Council. (2015b). *Climate Intervention: Reflecting Sunlight to Cool Earth*. The National Academies Press. https://doi.org/10.17226/18988

Nelson, J. A. (2008). Economists, value judgments, and climate change: A view from feminist economics. *Ecological Economics*, *65*(3), 441–447. https://doi.org/10.1016/j.ecolecon.2008.01.001

Nettleton, P. H. (2015). Brave Sperm and Demure Eggs: Fallopian Gender Politics on YouTube. *Feminist Formations*, *27*(1), 25–45.

Niemeier, U., Schmidt, H., Alterskjær, K., & Kristjánsson, J. E. (2013). Solar irradiance reduction via climate engineering: Impact of different techniques on the energy balance and the hydrological cycle. *J. Geophys. Res.*, *118*, 11905–11917. https://doi.org/10.1002/2013JD020445

Nordhaus, W. D. (2017). Revisiting the social cost of carbon. *Proceedings of the National Academy of Sciences*, *114*(7), 1518–1523. https://doi.org/10.1073/pnas.1609244114

Nowack, P. J., Abraham, N. L., Braesicke, P., & Pyle, J. A. (2016). Stratospheric ozone changes under solar geoengineering: Implications for UV exposure and air quality. *Atmospheric Chemistry and Physics*, *16*(6), 4191–4203. https://doi.org/10.5194/acp-16-4191-2016

Oldham, P., Szerszynski, B., Stilgoe, J., Brown, C., Eacott, B., & Yuille, A. (2014). Mapping the landscape of climate engineering. *Philosophical Transactions of the Royal Society A: Mathematical, Physical and Engineering Sciences*, *372*(2031), 20140065. https://doi.org/10.1098/rsta.2014.0065

Oomen, J. (2019). Anthropocenic Limitations to Climate Engineering. *Humanities*, *8*(4), 186. https://doi.org/10.3390/h8040186

Pawley, A., & Hoegh, J. (2011). Exploding Pipelines: Mythological Metaphors Structuring Diversity-Oriented Engineering Education Research Agendas. *2011 ASEE Annual Conference & Exposition Proceedings*, 22.684.1-22.684.21. https://doi.org/10.18260/1-2--17965

Pickering, A. (Ed.). (1992). *Science as practice and culture*. University of Chicago Press.

Pinto, I., Jack, C., Lennard, C., Tilmes, S., & Odoulami, R. C. (2020). Africa's Climate Response to Solar Radiation Management With Stratospheric Aerosol. *Geophysical Research Letters*, *47*(2), e2019GL086047. https://doi.org/10.1029/2019GL086047

Proctor, J. (2021). Atmospheric opacity has a nonlinear effect on global crop yields. *Nature Food*, *2*(3), 166–173. https://doi.org/10.1038/s43016-021-00240-w

Rahman, A. A., Artaxo, P., Asrat, A., & Parker, A. (2018). Developing countries must lead on solar geoengineering research. *Nature*, *556*(7699), 22–24. https://doi.org/10.1038/d41586-018-03917-8

Rasch, P. J., Tilmes, S., Turco, R. P., Robock, A., Oman, L., Chen, C.-C., Stenchikov, G. L., & Garcia, R. R. (2008). An overview of geoengineering of climate using stratospheric sulphate aerosols. *Phil. Trans. Roy. Soc. A*, *366*, 4007–4037. https://doi.org/10.1098/rsta.2008.0131

Ricke, K. L., Morgan, M. G., & Allen, M. R. (2010). Regional climate response to solar-radiation management. *Nature Geoscience*, *3*(8), 537–541. https://doi.org/10.1038/ngeo915

Rivers, D. L. (2019). Cartographies of Feminist Science Studies. *Women's Studies*, *48*(3), 177–185. https://doi.org/10.1080/00497878.2019.1603980

Salter, J., Robinson, J., & Wiek, A. (2010). Participatory methods of integrated assessment-a review: Participatory methods of integrated assessment. *Wiley Interdisciplinary Reviews: Climate Change*, *1*(5), 697–717. https://doi.org/10.1002/wcc.73

Sax, S. (2019, December 18). Geoengineering's Gender Problem Could Put the Planet at Risk. *Wired*. https://www.wired.com/story/geoengineerings-gender-problem-could-put-the-planet-at-risk/

Schneider, S. H. (1997). Integrated assessment modeling of global climate change: Transparent rational tool for policy making or opaque screen hiding value-laden assumptions? *Environmental Modeling and Assessment*, *2*(4), 229–249. https://doi.org/10.1023/A:1019090117643

Sedova, B., Kalkuhl, M., & Mendelsohn, R. (2020). Distributional Impacts of Weather and Climate in Rural India. *Economics of Disasters and Climate Change*, *4*(1), 5–44. https://doi.org/10.1007/s41885-019-00051-1

Shackley, S., Young, P., Parkinson, S., & Wynne, B. (1998). Uncertainty, Complexity and Concepts of Good Science in Climate Change Modelling: Are GCMs the Best Tools? *Climatic Change*, *38*(2), 159–205. https://doi.org/10.1023/A:1005310109968

Sheets, A. (2003). Has Feminism Changed Science? A Biological Perspective. *Kampf Writing Prize*. https://stuff.mit.edu/afs/athena.mit.edu/org/w/wgs/prize/as03.html

Shepherd, J., Caldeira, K., Cox, P., Haigh, J., Keith, D., Launder, B., Mace, G., MacKerron, G., Pyle, J., Rayner, S., Redgwell, C., & Watson, A. (2009). *Geoengineering the climate: Science, governance and uncertainty* (RS Policy document 10/09). The Royal Society.

Sikka, T. (2018). *Climate technology, gender, and justice: The standpoint of the vulnerable*. Springer.

Simonite, T. (2019, July 22). The Best Algorithms Struggle to Recognize Black Faces Equally. *Wired*. https://www.wired.com/story/best-algorithms-struggle-to-recognize-black-faces-equally/

Sovacool, B. K., Hess, D. J., Amir, S., Geels, F. W., Hirsh, R., Rodriguez Medina, L., Miller, C., Alvial Palavicino, C., Phadke, R., Ryghaug, M., Schot, J., Silvast, A., Stephens, J., Stirling, A., Turnheim, B., van der Vleuten, E., van Lente, H., & Yearley, S. (2020). Sociotechnical agendas: Reviewing future directions for energy and climate research. *Energy Research & Social Science*, *70*, 101617. https://doi.org/10.1016/j.erss.2020.101617

Stephens, J. C., & Surprise, K. (2020). The hidden injustices of advancing solar geoengineering research. *Global Sustainability*, *3*, e2. https://doi.org/10.1017/sus.2019.28

Suarez, P., & van Aalst, M. K. (2017). Geoengineering: A humanitarian concern: GEOENGINEERING: A HUMANITARIAN CONCERN. *Earth's Future*, *5*(2), 183–195. https://doi.org/10.1002/2016EF000464

Surprise, K. (2020). Geopolitical ecology of solar geoengineering: From a "logic of multilateralism" to logics of militarization. *Journal of Political Ecology*, *27*(1), 213–235. https://doi.org/10.2458/v27i1.23583

Tebaldi, C., Arblaster, J., & Knutti, R. (2011). Mapping model agreement on future climate projections: MAPPING MODEL AGREEMENT. *Geophysical Research Letters*, *38*(23), n/a-n/a. https://doi.org/10.1029/2011GL049863

Tilmes, S., Fasullo, J., Lamarque, J.-F., Marsh, D. R., Mills, M., Alterskjær, K., Muri, H., Kristjánsson, J. E., Boucher, O., Schulz, M., Cole, J. N. S., Curry, C. L., Jones, A., Haywood, J., Irvine, P. J., Ji, D., Moore, J. C., Karam, D. B., Kravitz, B., … Watanabe, S. (2013). The hydrological impact of geoengineering in the Geoengineering Model Intercomparison Project (GeoMIP). *J. Geophys. Res.*, *118*, 11036–11058. https://doi.org/10.1002/jgrd.50868

Trisos, C. H., Amatulli, G., Gurevitch, J., Robock, A., Xia, L., & Zambri, B. (2018). Potentially dangerous consequences for biodiversity of solar geoengineering implementation and termination. *Nature Ecology & Evolution*, *2*(3), 475–482. https://doi.org/10.1038/s41559-017-0431-0

Trogen, B. (2016). Aristotelian gender bias in modern depictions of fertilization. *Hektoen International: A Journal of Medical Humanities*, *Winter*. https://hekint.org/2017/01/29/aristotelian-gender-bias-in-modern-depictions-of-fertilization/

van der Sluijs, J. P. (2012). Uncertainty and Dissent in Climate Risk Assessment: A Post-Normal Perspective. *Nature and Culture*, *7*(2), 174–195. https://doi.org/10.3167/nc.2012.070204

Visioni, D., MacMartin, D. G., & Kravitz, B. (2021). Is Turning Down the Sun a Good Proxy for Stratospheric Sulfate Geoengineering? *Journal of Geophysical Research: Atmospheres*, *126*(5). https://doi.org/10.1029/2020JD033952

Weyant, J. (2017). Some Contributions of Integrated Assessment Models of Global Climate Change. *Review of Environmental Economics and Policy*, *11*(1), 115–137. https://doi.org/10.1093/reep/rew018

Wieding, J., Stubenrauch, J., & Ekardt, F. (2020). Human Rights and Precautionary Principle: Limits to Geoengineering, SRM, and IPCC Scenarios. *Sustainability*, *12*(21), 8858. https://doi.org/10.3390/su12218858

Wiertz, T. (2016). Visions of Climate Control: Solar Radiation Management in Climate Simulations. *Science, Technology, & Human Values*, *41*(3), 438–460. https://doi.org/10.1177/0162243915606524

Willey, A. (2016). A World of Materialisms: Postcolonial Feminist Science Studies and the New Natural. *Science, Technology, & Human Values*, *41*(6), 991–1014. https://doi.org/10.1177/0162243916658707

Winickoff, D. E., Flegal, J. A., & Asrat, A. (2015). Engaging the Global South on climate engineering research. *Nature Climate Change*, *5*(7), 627–634. https://doi.org/10.1038/nclimate2632

Wu, X., & Zhang, X. (2016). Automated Inference on Criminality using Face Images. *ArXiv*, 1611.04135.

Wylie, A. (2007). Doing Archaeology as a Feminist: Introduction. *Journal of Archaeological Method and Theory*, *14*(3), 209–216. https://doi.org/10.1007/s10816-007-9034-4

Xia, L., Robock, A., Cole, J., Curry, C. L., Ji, D., Jones, A., Kravitz, B., Moore, J. C., Muri, H., Niemeier, U., Singh, B., Tilmes, S., Watanabe, S., & Yoon, J.-H. (2014). Solar radiation management impacts on agriculture in China: A case study in the Geoengineering Model Intercomparison Project (GeoMIP). *Journal of Geophysical Research: Atmospheres*, *119*(14), 8695–8711. https://doi.org/10.1002/2013JD020630

Young, K., Fisher, J., & Kirkman, M. (2019). "Do mad people get endo or does endo make you mad?": Clinicians' discursive constructions of Medicine and women with endometriosis. *Feminism & Psychology*, *29*(3), 337–356. https://doi.org/10.1177/0959353518815704

Zarnetske, P. L., Gurevitch, J., Franklin, J., Groffman, P. M., Harrison, C. S., Hellmann, J. J., Hoffman, F. M., Kothari, S., Robock, A., Tilmes, S., Visioni, D., Wu, J., Xia, L., & Yang, C.-E. (2021). Potential ecological impacts of climate intervention by reflecting sunlight to cool Earth. *Proceedings of the National Academy of Sciences*, *118*(15), e1921854118. https://doi.org/10.1073/pnas.1921854118